\documentclass[12pt]{article}
\usepackage{cite,amssymb,amsmath}
\begin{document}
\title{Split Quaternions and Particles in (2+1)-Space}
\author{{\bf Merab Gogberashvili}\\
Andronikashvili Institute of Physics\\
6 Tamarashvili Street, Tbilisi 0177, Georgia\\
and\\
Javakhishvili State University\\
3 Chavchavadze Avenue, Tbilisi 0179, Georgia\\
{\sl (E-mail: gogber@gmail.com)}
}
\maketitle
\begin{abstract}
It is known that quaternions represent rotations in 3D Euclidean and Minkowski spaces. However, product by a quaternion gives rotation in two independent planes at once and to obtain single-plane rotations one has to apply by half-angle quaternions twice from the left and on the right (with its inverse). This 'double cover' property is potential problem in geometrical application of split quaternions, since (2+2)-signature of their norms should not be changed for each product. If split quaternions form proper algebraic structure for microphysics, representation of boosts in (2+1)-space leads to the interpretation of the scalar part of quaternions as wavelength of particles. Invariance of space-time intervals and some quantum behavior, like noncommutativity and fundamental spinor representation, probably also are algebraic properties. In our approach the Dirac equation represents the Cauchy-Riemann analyticity condition and the two fundamental physical parameters (speed of light and Planck's constant) appear from the requirement of positive definiteness of quaternionic norms.

\end{abstract}

\vskip 0.5cm


\section{Introduction}

Physics of (2+1)-space attracts considerable attention in different branches of physics, such as the theory of graphene \cite{Graphene}, black holes \cite{BH}, quantum gravity \cite{Carl}, AdS/CFT correspondence \cite{Wit} and gauge theory of gravity \cite{Chern-S}. It is known that rotations of 3D Euclidean and Minkowski spaces can be represented by the algebra of Hamilton's and split quaternions, respectively. Algebra of quaternions today are mainly used in the areas of computer graphics, navigation systems, to understand different aspects of physics and kinematics, and to rewrite in compact way some physical laws (see \cite{Computer,Qua} and references therein).

In general choice of convenient algebra should reveal some hidden properties of the physical system (as the solutions of adequate equations properly describe physical phenomena). The geometry of space-time cannot be described without the way it is observed, it is just reflections of symmetries of physical signals we receive and of the algebra using in the measurement process \cite{Go}. Since all observable quantities we extract from single measurements are real in geometrical applications it is possible to restrict ourselves to the field of real numbers. To have a transition from a manifold of the results of measurements to geometry, one must be able to introduce a distance between objects and an etalon for their comparison. Thus we need algebra with a unit element. In the algebraic language these physical requirements mean that to describe geometry we need normed composition algebra with the unit element over the field of real numbers. Besides of the usual real numbers, according to the Hurwitz theorem, there are three unique division algebras (complex numbers, quaternions and octonions) \cite{Sc}. Essential feature of all normed composition algebras is the existence of a real unit element and a different number of hyper-complex units. The square of the unit element is always positive while the squares of the hyper-complex units can be negative as well. In applications of division algebras mainly the elements with the negative square (similar to the complex unit), with the Euclidean norms, are used. Introducing the vector-like elements (with the positive squares) leads to split algebras having an equal number of terms with the positive and negative signs in the definition of their norms.

In this paper we consider applications of split quaternions for particle physics in 3D Minkowski space. We show that if split quaternions properly describe (2+1)-geometry, then some quantum characteristics of particles in this space can have algebraic roots.


\section{Quaternions}

Algebra of quaternions, discovered by William Hamilton in 1843, is an associative, non-commutative ring with four basis elements. However, its general element,
\begin{equation} \label{q}
q = q_0 + q_1i + q_2j + q_3 k = q_0 + q_1i + (q_2 + q_3 i)j ~,
\end{equation}
where $q_0, q_1, q_2, q_3$ are some real numbers, can be written by using only three basis elements - the unit element (which we denote as $1$) and the two hyper-complex units, $i$ and $j$, which anticommute with each other,
\begin{equation} \label{ij}
ij = - ji ~.
\end{equation}
The last hyper-complex basis element of quaternions, denoted by $k$ in standard Hamilton's notation, is possible to obtained by the product,
\begin{equation} \label{k}
k = (ij)~.
\end{equation}
According to the property (\ref{ij}) the basis unit (\ref{k}) also has anticommutative features with other two hyper-complex units:
\begin{equation}
i(ij) = - (ij)i ~, ~~~~~ j(ij) = - (ij)j~.
\end{equation}
Using the properties of the basis units under the conjugation,
\begin{equation} \label{*}
i^* = - i~, ~~~j^* = - j~, ~~~ (ij)^* = (j^*i^*) = - (ij) ~,
\end{equation}
the quaternion reverse to (\ref{q}), or the conjugated quaternion, can be constructed:
\begin{equation}
q^* = q_0 - q_1i - q_2j - q_3 (ij)~.
\end{equation}

Another useful representation of the quaternion (\ref{q}) is:
\begin{equation} \label{S+V}
q = S_q + V_q~, ~~~~~ q^* = S_q - V_q~,
\end{equation}
where the symbols
\begin{equation}
S_q = q_0~, ~~~~~ V_q = q_1i + q_2j + q_3 (ij)~,
\end{equation}
are called the scalar and vector parts of quaternion, respectively. In the representation (\ref{S+V}) composition of two quaternions, $q$ and $Q$, can be written in the form:
\begin{equation}\label{qQ}
qQ = S_qS_Q - (V_q \cdot V_Q)+ S_qV_Q + V_qS_Q - [V_q \times V_Q]~,
\end{equation}
where the products $(V_q \cdot V_Q)$ and $[V_q \times V_Q]$ mean the standard scalar and vector multiplications from vector algebra, respectively.

Note that the word 'vector' was first introduced by Hamilton just right for the pure imaginary part of quaternions, $V_q$ \cite{StNe}. He did not give geometrical meaning to the scalar part, $S_q$, but kept it to have division property of algebra. Later Hamilton used also the word 'versors' (meaning 'rotators') for the quaternionic hyper-complex basis units having negative square (that are similar to complex numbers), since he understood that such elements could be interpreted as representing rotation of vectors, instead of something that corresponds to a straight line in space. The algebra of Euclidean vectors was developed in 1880is by Gibbs and Heaviside from Hamilton's quaternions. They removed scalar part of quaternions and kept Hamilton's term 'vector' representing a pure quaternion, $V_q$. Since vectors have their origin in physical problems the definition of the products of vectors was obtained from the way in which such products occur in physical applications. Instead of the whole quaternion product (\ref{qQ}), which has a 'scalar' plus a 'vector' part, Gibbs and Heaviside defined two separate types of vector multiplications, ordinary scalar and vector products of Euclidean 3-vectors. They also changed the sign from minus to plus in the scalar product of quaternion algebra resulting to positive squares of all three hype-complex basis units. This changing corresponds to a shift of interpretation of basis units from 'versors' to unit polar vectors. It is known that Hamilton's pure quaternions are not equivalent to vectors \cite{SiMa}. For ordinary Euclidean vectors, basis elements are orthogonal unit polar vectors, while for Hamilton's quaternions they are unit 'versors' (imaginary units) and with respect to coordinate transformations behave similar to axial vectors. There are differences also in the proprieties of their products. The product of quaternions is associative, while both types of vector products are not.

As we already mentioned the main basis elements, $i$ and $j$, of Hamilton's quaternions are similar to standard complex unit, i.e. both are pure imaginary,
\begin{equation}
i^2 = j^2 = -1~.
\end{equation}
It turns out that properties of the third orthogonal basis unit of Hamilton's quaternions, $(ij)$, is analogous to $i$ and $j$, i.e.
\begin{equation}
(ij)(ij) = - i^2j^2 = -1~.
\end{equation}
Then for the norm of Hamilton's quaternions we have:
\begin{equation} \label{Nq+}
N = \sqrt {qq^*} = \sqrt {q^*q} = \sqrt {q_0^2 + q_1^2 + q_2^2 + q_3^2} ~,
\end{equation}
i.e. $N^2$ is positively defined.

When the two main basis units of quaternions shear the properties of unit polar vectors,
\begin{equation}
i^2= j^2 = 1~,
\end{equation}
we have the algebra of split quaternions. Now properties of the third basis element, $(ij)$, differs from the other two hyper-complex units since it again behaves like a pure imaginary object,
\begin{equation} \label{ij*}
(ij)(ij) = - i^2j^2 = -1 ~, ~~~~~(ij)^* = - (ij) ~.
\end{equation}
So $(ij)$ is similar to pseudo-vector and in geometrical applications it should not be considered as a real physical arrow in space, instead it finds satisfactory interpretation as a time-like dimension of (2+1)-Minkowski space.

Because of different properties of basis units the norm of a split quaternion,
\begin{equation} \label{Nq-}
N = \sqrt {q_0^2 - q_1^2 - q_2^2 + q_3^2}~,
\end{equation}
has (2+2)-signature and in general $N^2$ is not positively defined.

Note that one can justify the origins of complex numbers as a specific split quaternion,
\begin{equation} \label{z1}
z = x + (ij)y~,
\end{equation}
without introducing them ad-hoc. In this sense we can say that split quaternions are richer in structure compared to Hamilton's quaternions, since they contain complex numbers and Euclidean vectors as particular cases.


\section{Classification of split quaternions}

Algebra of split quaternions is an associative, non-commutative, non-division ring. One can define time-like, space-like and light-like split quaternions \cite{Oz-Er}:
\begin{eqnarray}
N^2 &<& 0 ~, ~~~~~ (space-like) \nonumber \\
N^2 &>& 0 ~, ~~~~~ (time-like) \\
N^2 &=& 0 ~, ~~~~~ (light-like)\nonumber
\end{eqnarray}
Space-like and time-like quaternions have multiplicative inverses,
\begin{equation}
q^{-1} = \frac {q^*}{N^2}~,
\end{equation}
with the property:
\begin{equation}
q q^{-1}= q^{-1} q = 1~,
\end{equation}
while light-like quaternions have no inverses. The vector part of any space-like quaternions is space-like but the vector part of any time-like quaternion can be space-like or time-like.

One can construct also polar forms of split quaternions \cite{Oz-Er}:
\begin{itemize}
\item {Every space-like quaternion can be written in the form:
\begin{equation}
q = N(\sinh \theta + \epsilon \cosh \theta)~,
\end{equation}
where
\begin{equation}
\sinh \theta = \frac {q_0}{N}~, ~~~~~ \cosh \theta = \frac {\sqrt{q_1^2 + q_2^2-q_3^2}}{N}~, ~~~~~\epsilon = \frac {q_1i + q_2j + q_3 (ij)}{\sqrt{q_1^2 + q_2^2-q_3^2}}~,
\end{equation}
and $\epsilon$ is a unit ($\epsilon^2 = 1$) space-like 3-vector.}
\item {Every time-like quaternion with the space-like vector part can be written in the form:
\begin{equation}
q = N(\cosh \theta + \epsilon \sinh \theta)~,
\end{equation}
where
\begin{equation}
\cosh \theta = \frac {q_0}{N}~, ~~~~~ \sinh \theta = \frac {\sqrt{q_1^2 + q_2^2-q_3^2}}{N}~, ~~~~~\epsilon = \frac {q_1i + q_2j + q_3 (ij)}{\sqrt{q_1^2 + q_2^2-q_3^2}}~,
\end{equation}
$\epsilon$ again is a unit space-like 3-vector.}
\item {Every time-like quaternion with the time-like vector part can be written in the form:
\begin{equation}
q = N(\cos \theta + \epsilon \sin \theta)~,
\end{equation}
where
\begin{equation}
\cos \theta = \frac {q_0}{N}~, ~~~~~ \sin \theta = \frac {\sqrt{-q_1^2 - q_2^2 + q_3^2}}{N}~, ~~~~~\epsilon = \frac {q_1i + q_2j + q_3 (ij)}{\sqrt{-q_1^2 - q_2^2 + q_3^2}}~,
\end{equation}
now $\epsilon$ is a unit time-like 3-vector.}
\end{itemize}

For $N \neq 0$ the quantity
\begin{equation}
\alpha = \frac {q}{N}
\end{equation}
is called unit split quaternion, which is useful to represent rotations of 3-Minkowski space.


\section{Zero divisors}

By the elements of split algebras special objects corresponding to the zeros of norms, called zero divisors, can be constructed \cite{Sc}. These critical elements of the algebra, which are similar to light-cone variables in Minkowski space-time, could serve as the unit signals characterizing physical events. The norms of split quaternions (\ref{Nq-}) have (2+2)-signature and we have two different types of 'light-cone'. Zero divisors corresponding to these cones are called projection operators and Grassmann numbers.

In the algebra of split quaternions two classes (totally four) projection operators,
\begin{equation} \label{Dq}
D^\pm_i = \frac{1}{2}(1 \pm i) ~, ~~~~~ D^\pm_j = \frac{1}{2}(1 \pm j)~,
\end{equation}
can be introduced. These two classes do not commute with each other. The commuting ones with the standard properties of projection operators:
\begin{equation} \label{D12}
D^+D^- = 0 ~, ~~~~~ D^\pm D^\pm = D^\pm~,
\end{equation}
are only the pairs $D^\pm_i$, or $ D^\pm_j$. The operators $D^+$ and $D^-$ differ from each other by the reflection of basis element and thus correspond to the direct and reverse critical signals along one of the two real directions, $i$ or $j$.

In the algebra we have also two classes of Grassmann numbers,
\begin{equation} \label{Gq}
G^\pm_i = \frac{1}{2}(1 \pm i)j ~, ~~~~~ G^\pm_j = \frac{1}{2}(1 \pm j)i ~,
\end{equation}
with the properties
\begin{equation}
G^\pm G^\pm = 0
\end{equation}
for each pair. The operators (\ref{Gq}) with different index do not commute with each other.

From the products of the zero divisors (\ref{Dq}) and (\ref{Gq}):
\begin{eqnarray} \label{DGq}
D^\pm_i G^\pm_i = G^\pm_i D^\mp_i = G^\pm_i ~, ~~~~~ D^\pm_i G^\mp_i = G^\pm_i D^\pm_i = 0 ~,\nonumber\\
D^\pm_j G^\pm_j = G^\pm_j D^\mp_j = G^\pm_j ~, ~~~~~ D^\pm_j G^\mp_j = G^\pm_j D^\pm_j = 0 ~,
\end{eqnarray}
we see that the Grassmann numbers and projection operators with the same indexes have pairwise commuting relations:
\begin{equation} \label{dg}
[D^\pm_i G^\mp_i] = 0 ~, ~~~~~ [D^\pm_j G^\mp_j] = 0 ~.
\end{equation}

Using commuting zero divisors any quaternion (\ref{q}) can be written in the form:
\begin{equation} \label{qDec}
q = D^+ \left(q_0 + q_1\right) + G^+ \left(q_2 + q_3 \right) + D^- \left(q_0 - q_1\right) + G^-\left(q_2 - q_3 \right) ~,
\end{equation}
where $D^\pm$ and $G^\pm$ are the projection operators and Grassmann elements belonged to the one of the classes from (\ref{Dq}) and (\ref{Gq}), labeled by $i$, or by $j$.

The quaternion algebra is associative and therefore can be represented by matrices. It is known that four basis elements of Hamilton's quaternions have the representation by the $2\times 2$ unit matrix and by the three ordinary complex Pauli matrices \cite{Qua}. Note that for a real matrix representation of Hamilton's quaternions one needs 4-dimensional matrices.

In the case of split quaternion we get the simplest non-trivial representation by the three $2\times 2$ real Pauli matrices accompanied by the unit matrix. Namely, the main two basis elements of split quaternions have the following matrix representation:
\begin{equation}
i =
\begin{pmatrix}
1 & 0 \cr
0 & -1
\end{pmatrix}~, ~~~~~
j =
\begin{pmatrix}
0 & 1 \cr
1 & 0
\end{pmatrix}~,
\end{equation}
and the third unit basis element $(ij)$, which is formed by multiplication of $i$ and $j$, has the representation:
\begin{equation}
(ij) =
\begin{pmatrix}
0 & 1 \cr
-1 & 0
\end{pmatrix}~.
\end{equation}
In contrast with the ordinary Pauli matrices, the squares of the real Pauli matrices give the unit matrix with the different signs,
\begin{equation} \label{square}
i^2 = j^2 = (1) ~, ~~~~~(ij)^2 = -(1) ~.
\end{equation}

Conjugation of the unit elements means changing of signs of corresponding matrices and
\begin{equation} \label{i-norm}
ii^* = jj^* = -(1) ~, ~~~~~(ij)(ij)^* = (1) ~.
\end{equation}

Matrix representation of the independent projection operators and Grassmann elements from (\ref{Dq}) and (\ref{Gq}) labeled by $i$ are:
\begin{eqnarray}
D^+_i = \frac{1}{2}(1+i) =
\begin{pmatrix}
1 & 0\cr
0 & 0
\end{pmatrix}~, ~~~~~
D^-_i =\frac{1}{2}(1-i) =
\begin{pmatrix}
0 & 0\cr
0 & 1
\end{pmatrix}~,\nonumber
\\
G^+_i =\frac{1}{2}(j+ij) =
\begin{pmatrix}
0 & 1\cr
0 & 0
\end{pmatrix}~, ~~~~~
G^-_i = \frac{1}{2}(j-ij) =
\begin{pmatrix}
0 & 0\cr
1 & 0
\end{pmatrix}~.
\end{eqnarray}
It is easy to find matrix representation of zero divisors labeled by the second index $j$ also.

The decomposition (\ref{qDec}) of a split quaternion now can be written in the form:
\begin{equation}
q =
\begin{pmatrix}
\left(q_0 + q_1\right) & \left(q_2 + q_3 \right) \cr
\left(q_2-q_3\right) & \left(q_0 - q_2\right)
\end{pmatrix}
\end{equation}
and the norm (\ref{Nq-}) is constructed by the determinant of this matrix,
\begin{equation} \label{q-norm}
{\it det} q = \left(q_0^2 - q_1^2\right) - \left(q_2^2 - q_3^2\right) ~.
\end{equation}


\section{Rotations}

There are a lot of methods to represent rotations in Euclidean 3-space, like orthonormal matrices and Euler angles, however, Hamilton's quaternions are the most convenient ones \cite{Qua}. If we compare to orthonormal matrices, there are some constraints as each colon of an orthonormal matrix must be unit vector and must be perpendicular to each other. These constraints make it difficult to construct an orthonormal matrix using nine numbers. But, we can construct easily a rotation orthonormal matrix using a unit quaternion,
\begin{equation} \label{alpha}
\alpha = \frac qN = \cos \frac {\theta}{2} + \sin \frac {\theta}{2}\epsilon~,
\end{equation}
where $\epsilon$ is a unit pseudo-vector like object ($\epsilon^* = - \epsilon$ and the inner product $\epsilon^2 = -1$) having three components. The vector part of (\ref{alpha}) can be thought of as a vector about which rotation should be performed and scalar part specifies the amount of rotation that should be performed about the vector part. That is, only four numbers are enough to represent a rotation by quaternions and there is only one constraint - the unity of norm. This makes it possible to find solutions to some optimization problems involving rotations, which are hard to solve when using orthonormal matrices because of the six non-linear constraints to enforce orthonormality, and the additional constraint - the unity of determinant.

Every unit quaternion (\ref{alpha}) represents a rotation in the Euclidean 3-space by two side multiplication. Result of the product
\begin{equation} \label{aqa}
q'= \alpha q \alpha ^*
\end{equation}
is the quaternion $q'$ whose norm and scalar part are the same as for $q$ and the vector part $V_{q'}$ is obtained by revolving $V_{q}$ conically about $ \epsilon$ through the angle $\theta$. For example, we can represent the rotations (\ref{aqa}) about the standard coordinate axes $x$, $y$ and $z$ with the unit quaternions:
\begin{equation}
\alpha_x = \left(\cos \frac {\theta}{2}, \sin \frac {\theta}{2}, 0, 0 \right)~, ~~~
\alpha_y = \left(\cos \frac {\theta}{2}, 0, \sin \frac {\theta}{2}, 0 \right)~, ~~~
\alpha_z = \left(\cos \frac {\theta}{2}, 0, 0, \sin \frac {\theta}{2} \right)~.
\end{equation}

The set of unit quaternions (\ref{alpha}) form a group that is isomorphic to SU(2) and is a double cover of SO(3), the group of 3-rotations. Under these isomorphisms the quaternion multiplication operation corresponds to the composition operation of rotations.

But why quaternions have the 'double-cover' property (\ref{aqa}), why there are two different quaternions ($\alpha$ and $\alpha^{-1} = \alpha^*$, negatives of each other) that represent the same 3D rotation instead of $\alpha (\theta) q$ in analogy with complex numbers? It turns out that multiplication by a single quaternion do represent rotations in 4D quaternionic space and not in 3D. In the case of complex numbers we have 2D space and just one dimension of rotation. In 3D, we talk about rotating about an axis, but we really mean rotating in a plane perpendicular to that axis. In 4D space of quaternions there are enough dimensions that it's possible to rotate in two independent planes at once. The planes have no axes in common; they intersect only at a single point, which is the center of rotation, so both rotations can take place without disturbing each other (not possible in 3D, where two planes always intersect in a line). For example, consider the rotation of a quaternion $q$ around the axis $z$ by the unit quaternion
\begin{equation}
\alpha_z = \cos \frac {\theta}{2} + \sin \frac {\theta}{2}(ij)~.
\end{equation}
The result of the left multiplication,
\begin{eqnarray}
q' = \alpha_z q &=& \left(\cos \frac {\theta}{2} q_0 -\sin \frac {\theta}{2} q_3\right) + \left(\cos \frac {\theta}{2} q_1 +\sin \frac {\theta}{2} q_2\right)i + \nonumber \\
 &+& \left(\cos \frac {\theta}{2} q_2 - \sin \frac {\theta}{2} q_1\right)j+ \left(\cos \frac {\theta}{2} q_3 + \sin \frac {\theta}{2} q_0\right) (ij) ~,
\end{eqnarray}
gives simultaneous rotations in $(q_0 - q_3)$ and $(q_1 - q_2)$ planes by the same angle $\theta/2$.

So a quaternion always rotates in two independent planes at once, by the same angle. Moreover, one of the two planes always includes the axis of real numbers $q_0$. This is not what we want for 3D rotation; we need to be able to rotate in just one Euclidean plane. It turns out that swapping the order of multiplication of two quaternions, i.e. the right product $q\alpha$, will reverse the direction of rotation in one of the two planes - namely, the one that does not contain the axis of real numbers. To get a rotation that only affects a single plane in 4D, we have to apply the quaternion twice, multiplying on both the left and on the right (with its inverse), as in (\ref{aqa}). The rotation we want gets done twice (and the one we don't want gets canceled out), so we have to halve the angle $\theta$ going in to make up for it.

So the elementary rotations by unit Hamilton's quaternions are represented by half angles and full rotations does not affect the scalar part $q_0$. In general geometrical interpretation of the scalar part of quaternions caused difficulty, Hamilton himself tried without notable success to interpret it as an extra-spatial unit \cite{StNe}. Later when Gibbs and Heaviside introduce vector algebra they even removed at all the scalar part of quaternions. Introduction of vectors in physics was successful, however because of removing of the scalar part of quaternions the division operation is not defined for vectors. There was lost also the property of 'versors', that they are rotation generators and expresses not only the final state achieved after a rotation, but the direction in which this rotation has been performed. It is this direction of rotation that the standard matrix representation of the rotation group fails to give.


\section{Boosts}

The vector part of split quaternions, instead of the Euclidean 3-space, represents the (2+1)-Minkowski space-time \cite{Oz-Er}. Now we want to consider rotations performed by the unit split quaternions in a similar manner as Hamilton's unit quaternions do.

Let us define the interval in (2+1)-Minkowski space by the element of split quaternions \cite{Go},
\begin{equation} \label{s}
s = \lambda + xi + yj + ct (ij) ~,
\end{equation}
where the four real parameters that multiply the basis units denotes: some quantity $\lambda$ with the dimensions of length (whose physical meaning we shall discuss below); the spatial coordinates $x$ and $y$; and the time coordinate $t$, accompanied by the fundamental constant of the speed of light $c$. Using the conjugation rules of split quaternion basis units one can find that the norm of (\ref{s}), the interval,
\begin{equation} \label{sN}
s^2 = s s^* = s^* s = \lambda^2 - x^2 - y^2 + c^2t^2 ~,
\end{equation}
has (2+2)-signature and in general is not positively defined. As in the standard relativity we require
\begin{equation} \label{s>0}
s^2 \geq 0~.
\end{equation}

As it was shown in \cite{Oz-Er} rotations in the 3-Minkowski space can be generated only with unit time-like split quaternions, since the set of space-like quaternions do not form a group (it is not closed under multiplication). Whereas, the set of time-like quaternions forms a group under the split quaternion product, automorphism group of split quaternions, which is isomorphic to $SO(2,1)$.

If the vector part of the split quaternion is time-like/space-like, then the rotation angle $\theta$ is spherical/hyperbolic. Correspondingly, the pair of unit time-like quaternions with respect to the time-like, or space-like, axis of rotation $\epsilon$ are $\pm (\cos \theta/2 + \epsilon\sin \theta /2 )$, or $\pm (\cosh \theta/2 + \epsilon\sinh \theta /2 )$, respectively. For example, suppose we rotate a split quaternion (\ref{s}) around the $t$-axis. The axis is time-like and we must use time-like unit quaternion (which is analogous to ordinary complex numbers) with time-like vector part and spherical rotation angle,
\begin{equation} \label{Rq}
\alpha_t = \cos \frac {\theta}{2} \pm \sin \frac {\theta}{2}(ij) ~.
\end{equation}
The left product, $\alpha_ts$, now represents the simultaneous rotations by the angle $\theta /2$ in two orthogonal planes - $(\lambda -t)$ and $(x-y)$. Acting from both side on $s$ by $\alpha_t $ and $\alpha_t^* $, the rotation in $(\lambda -t)$-plane cancels out and we left with the rotation in the real $(x-y)$-plane by the angle $\theta$. These transformations form the group $O(2)$.

Now let us consider boosts. Rotations of $s$ around the space-like $y$-axis, or boosts along $x$ by the velocity $v_x$, can be done using the time-like unit split quaternion with the space-like vector part,
\begin{equation}
\alpha_y = \cosh \frac\phi2 - \sinh \frac\phi2 j ~,
\end{equation}
where the hyperbolic angle $\phi$ relates to the velocity by the standard relativistic expressions,
\begin{equation}
\cosh \phi = \frac {1}{\sqrt{1 - v_x^2/c^2}} ~, ~~~~~ \sinh \phi = \frac {v_x}{c} \cosh \phi~.
\end{equation}
One-side transformation of $q$, or the result of left multiplication has the form:
\begin{equation}
q' = \alpha q = \lambda' + t' ij + y' j + x' i ~,
\end{equation}
were
\begin{eqnarray} \label{'}
\lambda' &=& \cosh \frac\phi2 \lambda - \sinh \frac\phi2 y ~,\nonumber \\
t' &=& \cosh \frac\phi2 t + \sinh \frac\phi2 x~, \nonumber \\
y' &=& \cosh \frac\phi2 y - \sinh \frac\phi2 \lambda ~, \\
x' &=& \cosh \frac\phi2 x + \sinh \frac\phi2 t ~. \nonumber
\end{eqnarray}
Acting from both side on $s$ by $\alpha_y $ and $\alpha_y^* $, the rotation in $(\lambda -y)$-plane cancels out and we left with the one parameter group of boosts in $(t-x)$-plane by the angle $\phi$.

However, there are some peculiarities due to the double-cone structure of the norms of split quaternions (\ref{sN}) and we should be sure that the condition of positivity of the norms (\ref{s>0}) comes true for each left and right rotations. This problem is absent for Hamilton's quaternions, because of Euclidean structure of the norms (\ref{Nq+}), and is hidden for split quaternions due to the double-cover rotation property.

Analysis of quaternionic boosts can help us in physical interpretation of the scalar part $\lambda$ of (\ref{s}). From (\ref{'}) we notice that for the boosts along the positive $x$-direction, i.e. when $x'$ increases (or the $x$-component of the momentum, $p_x$, increases), the quantity $\lambda'$ decreases and vice versa. In quantum mechanics the quantity with the dimension of length which is inversely proportional to the momentum is particle's wavelength. So in geometrical application it is natural to interpret the scalar part of split quaternion (\ref{s}) as the wavelength describing the inertial properties of particle's reference frame.

Suppose in a laboratory coordinate system the particle with the wavelength $\lambda$ moves along the $x$-axis with the velocity $v_x$. The wavelength in particles own system we denote by $\Lambda$. Then from the condition of invariance of the intervals,
\begin{equation} \label{ds-inv}
ds = d\lambda \left(1 + j \frac {dy}{d\lambda}\right) + (ij) ct\left( 1 + j \frac {v_x}{c} \right) = d\Lambda + c d\tau~,
\end{equation}
where $\tau$ is the proper time of the particle, we obtain the two conditions:
\begin{equation} \label{c}
\left|\frac{d\tau}{dt}\right| = \sqrt {1 - \frac {v_x^2}{c^2}}~, ~~~~~ \left |\frac{d\Lambda}{d\lambda}\right| = \sqrt {1 - \frac {dy^2}{d\lambda^2}}~.
\end{equation}
From these conditions follows the relations:
\begin{equation} \label{frac}
\left|\frac{v_x}{c}\right| < 1 ~, ~~~~~ \left |\frac{d y}{d \lambda}\right| < 1 ~,
\end{equation}
which must be obeyed simultaneously. This means that, together with $c$, there must exist the second fundamental constant (which can be extracted from $\lambda$) characterizing this critical property of the algebra.

In (2+2)-space there are two different light-cones, and we have two classes of zero divisors, projection operators and Grasmann numbers, corresponding to the critical rotations of these cones and there must exist two fundamental constants characterizing this property of algebra. We know that the parameter with required properties is the Planck constant $\hbar$, which relates particles wavelength to its momentum,
\begin{equation}
\lambda = \frac {\hbar}{p}~.
\end{equation}
So in our approach two fundamental physical constants, $c$ and $\hbar$, have the geometrical origin and correspond to two kinds of critical signals in (2+2)-space.

Also we have seen that for the boosts with the positive velocity, when $p_x$ increases ($p_y$ decreases), the quantity $\lambda$ decreases, i.e. $d\lambda < d\Lambda$, and vice versa. So for the change in the scalar part of the split quaternion, $\Delta \lambda$, when the particle momentum in $x$-direction increses, we can write
\begin{equation} \label{P}
\Delta \lambda \sim - \frac{\hbar}{\Delta p_y} ~,
\end{equation}
were $p_y$ is the $y$-component of the momentum. When the variety of wave lengths becomes overall shorter, the overall magnitude of the variety of momenta must become greater, i.e. the shorter the wavelength, the higher will be its frequency and hence carry a greater amount of momentum. Similar relation can be obtained for the boost along the $y$-axis.

Inserting (\ref{P}) into (\ref{frac}) we can conclude that the uncertainty principle,
\begin{equation} \label{U}
\Delta x \Delta p_x \gtrsim \hbar ~,
\end{equation}
probably has the same geometrical meaning as the existence of the maximal velocity.


\section{Quaternionic Dirac equation}

There exist several successful applications of quaternions (see, for example \cite{QD}) in formulating of the Dirac equation. In this purpose it is important to define the gradient operator for quaternion-valued functions. Although there have been some derivations of this operator in literature with different level of details (for example, see \cite{d} and references therein), it is still not fully clear how this operator can be written in the most general case and how it can be applied to various functions.

Using analogy with complex analysis for (\ref{s}) we can define the split quaternionic derivative operator,
\begin{equation} \label{d}
\frac {d}{ds} = \frac 12 \left[ \frac {d}{d\lambda} + i \frac {d}{dx} + j \frac {d}{dy} + (ij)\frac {d}{cdt}\right]~,
\end{equation}
such that its action upon $s$ is:
\begin{equation}
\frac {ds}{ds} = 1~,
\end{equation}
while it gives zero if applied to
\begin{equation}
s^* =  \lambda - xi - yj - ct (ij)~.
\end{equation}
Similarly, the conjugated derivative operator can be defined by the operator
\begin{equation} \label{d*}
\frac {d}{ds^*} = \frac 12 \left[ \frac {d}{d\lambda} - i \frac {d}{dx} - j \frac {d}{dy} - (ij)\frac {d}{cdt}\right]~,
\end{equation}
which annihilates $s$. Thus from the definitions of quaternionic gradients, (\ref{d}) and (\ref{d*}), we find
\begin{equation}
\frac {ds^*}{ds} = \frac {ds}{ds^*} = 0~.
\end{equation}
From these relations it is clear that the interval (\ref{sN}) is a constant function for the restricted left quaternionic gradient operators,
\begin{eqnarray}
\frac {d}{ds_L} \left(s^* s\right) = \left(\frac {ds^*}{ds} \right) s = 0~, \nonumber \\
\frac {d}{ds^*_L} \left(ss^*\right) = \left(\frac {ds}{ds^*} \right) s^* = 0~.
\end{eqnarray}
So invariance of intervals in the space of split quaternions can be an algebraic property as well.

To define derivative of a split quaternionic function,
\begin{equation} \label{Phi}
\Phi (s,s^*) = \phi_\lambda + i\phi_x +i \phi_y + (ij) \phi_t~,
\end{equation}
we need the condition of analyticity of functions of a split quaternion variable,
\begin{equation} \label{dphi/ds}
\frac {d\Phi(s,s^*)}{ds^*} = 0~,
\end{equation}
which is similar to the Cauchy-Riemann equations from complex analysis. This statement, that quaternionic functions should be independent of the variable $s^*$, represents the condition that quaternionic derivative be independent of direction along which it is evaluated \cite{Ana}.

The algebraic condition (\ref{dphi/ds}) can be understood as the quaternionic Dirac equations in (2+1)-space. Indeed, for the real-valued functions $\phi_\nu$ ($\nu = \lambda, x, y, t$), the condition (\ref{dphi/ds}) is equivalent to the system of equations:
\begin{eqnarray} \label{phi}
\frac {\partial \phi_\lambda}{d \lambda} - \frac {\partial \phi_x}{d x} - \frac {\partial \phi_y}{d y} + \frac {\partial \phi_t}{d t} = 0~, \nonumber \\
\frac {\partial \phi_x}{d \lambda} - \frac {\partial \phi_\lambda}{d x} + \frac {\partial \phi_t}{d y} - \frac {\partial \phi_y}{d t} =0~, \nonumber \\
\frac {\partial \phi_y}{d \lambda} - \frac {\partial \phi_t}{d x} - \frac {\partial \phi_\lambda}{d y} + \frac {\partial \phi_x}{d t} =0~, \\
\frac {\partial \phi_t}{d \lambda} - \frac {\partial \phi_y}{d x} + \frac {\partial \phi_x}{d y} - \frac {\partial \phi_\lambda}{d t} = 0~. \nonumber
\end{eqnarray}
Since in this system the terms with $\lambda$-derivatives has the same sign we can separate the variables in the form:
\begin{equation} \label{phi=psi}
\phi_\nu (\lambda,x,y,t) = e^{-m\lambda}\psi_\nu (t,x,y)~, ~~~~~(\nu = \lambda, x, y, t)
\end{equation}
where $m$ is the mass parameter ($\lambda$ defines inertial properties of particles). Then (\ref{phi}) reduces to the system:
\begin{eqnarray} \label{psi}
- \frac {\partial \psi_y}{d x} + \frac {\partial \psi_x}{d y} - \frac {\partial \psi_\lambda}{d t}  + m \psi_t= 0~, \nonumber \\
\frac {\partial \psi_x}{d x} + \frac {\partial \psi_y}{d y} - \frac {\partial \psi_t}{d t} - m \psi_\lambda = 0~, \nonumber \\
\frac {\partial \psi_t}{d x} + \frac {\partial \psi_\lambda}{d y} - \frac {\partial \psi_x}{d t} - m \psi_y = 0~, \\
- \frac {\partial \psi_\lambda}{d x} + \frac {\partial \psi_t}{d y} - \frac {\partial \psi_y}{d t} + m \psi_x = 0~. \nonumber
\end{eqnarray}
From the other hand this system is equivalent to the (2+1)-dimensional free Dirac equation for a massive particle,
\begin{equation} \label{Dirac}
\left(i\gamma^\alpha\partial_\alpha - m\right) \Psi (t,x,y) = 0~, ~~~~~(\alpha = 0, 1, 2)
\end{equation}
where $i$ now denotes standard complex unit and we use the system of units where $c = \hbar = 1$. For the gamma-matrices we choice the representation:
\begin{equation}
\gamma^0 = \sigma^3 =
\begin{pmatrix}
1 & 0\cr
0 & -1
\end{pmatrix}~, ~~~~~
\gamma^1 = \sigma^1 =
\begin{pmatrix}
0 & i\cr
i & 0
\end{pmatrix}~, ~~~~~
\gamma^2 = i\sigma^2 =
\begin{pmatrix}
0 & 1\cr
-1 & 0
\end{pmatrix}~.
\end{equation}
If four components of the complex spinor $\Psi$ in (\ref{Dirac}) we identify with the components of the split quaternion (\ref{phi=psi}) in the form:
\begin{equation} \label{Psi}
\Psi (t,x,y) =
\begin{pmatrix}
-\psi_\lambda - i \psi_t \cr
\psi_x + i \psi_y
\end{pmatrix}~,
\end{equation}
then the quaternionic Cauchy-Riemann conditions (\ref{dphi/ds}) become equivalent to the complex (2+1)-Dirac equation (\ref{Dirac}).

The free particle solution of the Dirac equation (\ref{Dirac}), with the norming constant equal to unity, is easy to find:
\begin{equation}
\Psi (t,x,y) =
\begin{pmatrix}
- \frac {p_y+ip_x}{E+m} \cr
1
\end{pmatrix} e^{i \left(Et + p_xx + p_yy \right)}~.
\end{equation}
Then from (\ref{Psi}) we find the coefficients of the split quaternion (\ref{Phi}) that solve quaternionic Cauchy-Riemann condition (\ref{dphi/ds}) and describe quaternionic potential flow:
\begin{eqnarray}
\phi_\lambda &=& \frac {p_y\cos \left( Et + p_xx + p_yy\right)  - p_x \sin \left( Et + p_xx + p_yy\right)}{E+m}~e^{-m \lambda}~, \nonumber \\
\phi_x &=& \cos \left( Et + p_xx + p_yy\right) e^{-m \lambda}~, \nonumber \\
\phi_y &=& \sin \left( Et + p_xx + p_yy\right) e^{-m \lambda}~,  \\
\phi_t &=& \frac {p_x \cos \left( Et + p_xx + p_yy\right) + p_y \sin \left( Et + p_xx + p_yy\right)}{E+m}~e^{-m \lambda}~. \nonumber
\end{eqnarray}


\section{Conclusions}

Connection of the algebra of split quaternions with the quantum properties of particles in (2+1)-space were considered. There exists well known problem of interpretation of scalar part of quaternions, only a few attempts in this direction were made. For instance, in color image representations scalar part of quaternions was successful interpreted as the reference color \cite{Ang}. In our opinion in application to microphysics the scalar part of split quaternions can govern inertial properties of the reference frames of particles and can be interpreted as the wavelength. It is known that physical limits on masses of reference systems cause large uncertainties in the measurement even for classical objects \cite{NgDa}. We think that invariance of space-time interval and some quantum behaviors, like noncommutativity and spinor representations with the half-angle rotation properties, are encoded in the properties of quaternions. In our approach two fundamental physical constants (light speed and Planck's constant) have similar geometrical meanings and appear from the positive definiteness of quaternionic norms and the Dirac equation appears to represent the condition of quaternionic analyticity, analog of the Cauchy-Riemann equations from complex analysis.


\section*{Acknowledgments:}

The research was supported by the grant of Shota Rustaveli National Science Foundation $\#{\rm GNSF/ST}09\_798\_4-100$.


\end{document}